\documentclass[prl,preprint,showpacs,superscriptaddress]{revtex4}

\usepackage{graphicx}
\usepackage{dcolumn}
\usepackage{bm}

\usepackage{color}

\begin{document}


\title{Surface Tomonaga-Luttinger liquid state on Bi/InSb(001)}

\author{Yoshiyuki Ohtsubo}
\email{y_oh@fbs.osaka-u.ac.jp}
\affiliation{Graduate School of Frontier Biosciences, Osaka University, Suita 565-0871, Japan}
\affiliation{Department of Physics, Graduate School of Science, Osaka Unviersity, Toyonaka 560-0043, Japan}
\affiliation{Synchrotron SOLEIL, Saint-Aubin-BP 48, F-91192 Gif sur Yvette, France}
\author{Jun-ichiro Kishi}
\author{Kenta Hagiwara}
\affiliation{Department of Physics, Graduate School of Science, Osaka Unviersity, Toyonaka 560-0043, Japan}
\author{Patrick Le F\`evre}
\author{Fran\c{c}ois Bertran}
\affiliation{Synchrotron SOLEIL, Saint-Aubin-BP 48, F-91192 Gif sur Yvette, France}
\author{Amina Taleb-Ibrahimi}
\affiliation{Synchrotron SOLEIL, Saint-Aubin-BP 48, F-91192 Gif sur Yvette, France}
\affiliation{UR1/CNRS-Synchrotron SOLEIL, Saint-Aubin, F-91192 Gif sur Yvette,
France}
\author{Hiroyuki Yamane}
\author{Shin-ichiro Ideta}
\author{Masaharu Matsunami}
\author{Kiyohisa Tanaka}
\affiliation{Institute for Molecular Science, Okazaki 444-8585, Japan}
\author{Shin-ichi Kimura}
\email{kimura@fbs.osaka-u.ac.jp}
\affiliation{Graduate School of Frontier Biosciences, Osaka University, Suita 565-0871, Japan}
\affiliation{Department of Physics, Graduate School of Science, Osaka Unviersity, Toyonaka 560-0043, Japan}

\date{\today}

\begin{abstract}
A 1D metallic surface state was created on an anisotropic InSb(001) surface covered with Bi. Angle-resolved photoelectron spectroscopy (ARPES) showed a 1D Fermi contour with almost no 2D distortion.
Close to the Fermi level ($E_{\rm F}$), the angle-integrated photoelectron spectra showed power-law scaling with the binding energy and temperature.
The ARPES plot above $E_{\rm F}$ obtained thanks to thermally broadened Fermi edge at room temperature showed a 1D state with continuous metallic dispersion across $E_{\rm F}$ and power-law intensity suppression around $E_{\rm F}$.
These results strongly suggest a Tomonaga-Luttinger liquid on the Bi/InSb(001) surface. 
\end{abstract}

\maketitle

The Fermi liquid theory of ordinal three-dimensional (3D) metals breaks down in one-dimensional (1D) systems to produce various exotic quantum phases.
Tomonaga-Luttinger liquid (TLL) \cite{Tomonaga50, Luttinger63} is an exactly solvable model of a gapless 1D quantum system, that is characterized by power-law scaling and spin-charge separation for low-energy excitation spectra \cite{Voit94}.
So far, various 1D systems have been studied as TLL candidates.
Only a few have shown metallic states with power-law spectral features, e.g., carbon nanotubes \cite{Ishii03} and lithium purple bronze \cite{Wang06, Dudy13}.
The other characteristic of spin-charge separation has been reported in some 3D materials with 1D electronic structures \cite{Claessen02, Kim06, Jompol09}.

The surface of semiconductors is known to show various self-assembled 1D atomic structures that are regarded as suitable systems for studying 1D metallic states \cite{Himpsel01, Grioni09}, such as In/Si(111) \cite{Nogami87}, Au/Si(557) \cite{Segovia99}, and Pt/Ge(001) \cite{Gurlu03}.
On such surface systems, $in$-$situ$ electron/hole doping can be performed by the deposition of additional atoms, and the local atomic structure can be observed/controlled by scanning probe techniques.
These manipulations on the surface 1D states can provide further insight into 1D physics. Hence, surface TLLs have attracted much attention in the past two decades.
However, despite the various 1D surfaces, most of them do not behave as TLLs at low temperature. 
Possible reasons for the non-TLL behavior have been discussed for each specific system, such as a metal-insulator transition at low temperatures \cite{Yeom99} and interaction with other two-dimensional (2D) surface bands \cite{Yaji13}.

So far, the only candidate for a surface TLL is an Au nanowire assembled on Ge(001) (Au/Ge(001)), which shows power-law scaling for low-energy spectral features according to scanning tunneling spectroscopy (STS) and angle-resolved photoelectron spectroscopy (ARPES) \cite{Blumenstein11, Meyer14}.
However, a recent work presented doubt with regard to the TLL picture and proposed an alternative interpretation of the STS spectra: two different and gapped states lying just above/below the Fermi level ($E_{\rm F}$) \cite{Park14}.
This picture can also explain the STS spectral shape around $E_{\rm F}$.
In order to distinguish these two possibilities without ambiguity, the 1D band dispersion needs to be traced across $E_{\rm F}$ continuously with $k$ resolution in reciprocal space. 
However, to the best of our knowledge, no such experimental data have been reported about surface TLLs so far.
While some ARPES results have been given for Au/Ge(001) \cite{Blumenstein11, Meyer14, Nakatsuji11}, the dispersion above $E_{\rm F}$ is not accessible with these data.

In this letter, we report on the case of a new surface TLL candidate discovered with ARPES on a Bi-induced anisotropic structure on InSb(001) (Bi/InSb(001)).
The Bi/InSb(001) surface state showed a 1D Fermi contour (FC) with almost no undulation and stayed metallic down to 35 K.
The 1D feature is the unique contour on $E_{\rm F}$ of the sample, which showed a parabolic dispersion with its bottom at the center of the surface Brillouin zone (SBZ).
The photoelectron spectra in the proximity of $E_{\rm F}$ obey power-law scaling as a function of binding energy and sample temperature with the same power index $\alpha$ = 0.7 $\pm$0.1.
The ARPES plot divided by the Fermi distribution function at room temperature (RT) indicated the 1D state with continuous metallic dispersion across $E_{\rm F}$ and power-law intensity suppression around $E_{\rm F}$.
These results strongly suggest that the 1D surface structure of Bi/InSb(001) hosts a surface TLL.

\begin{figure}[p]
\includegraphics[width=80mm]{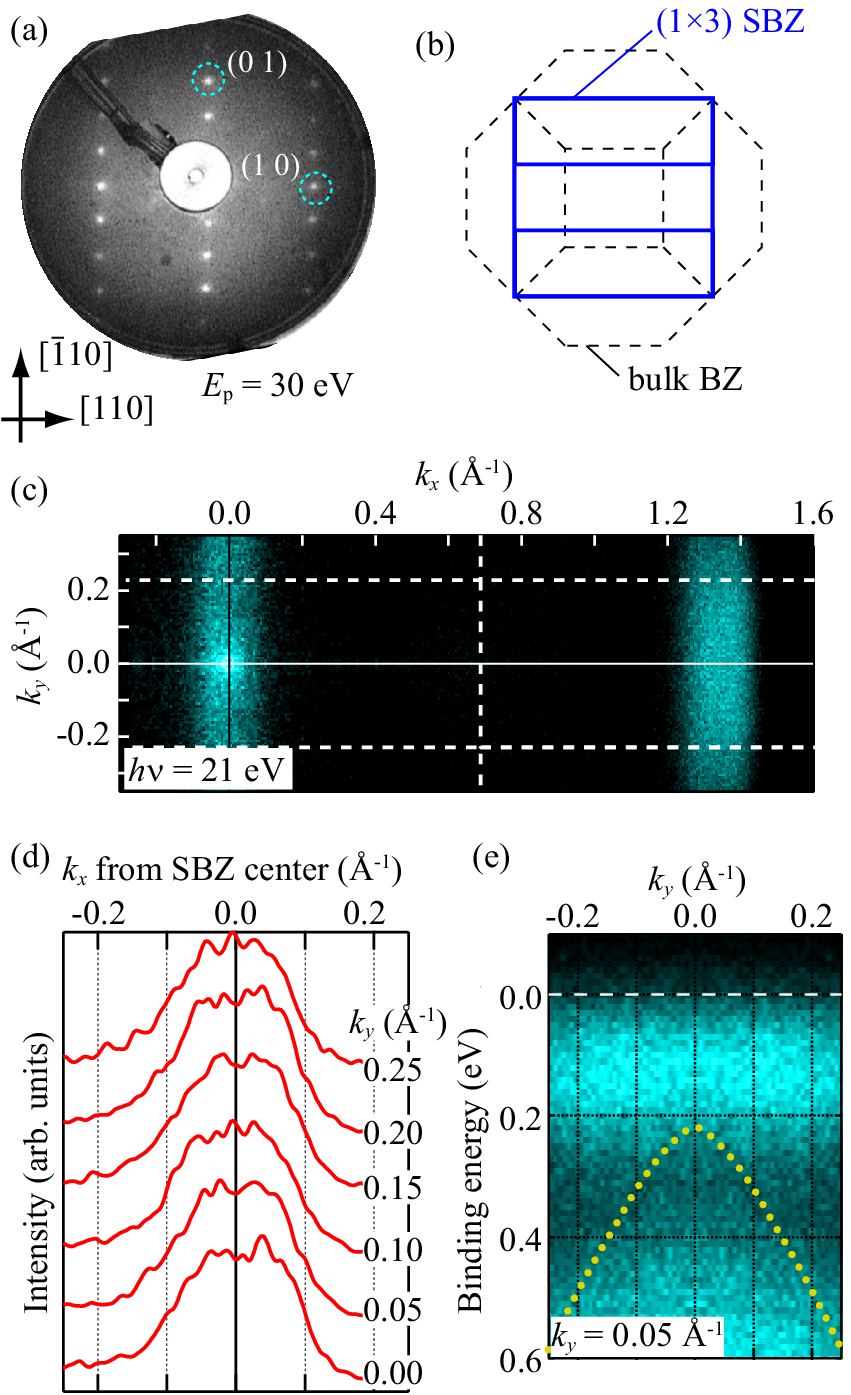}
\caption{\label{fig1} (Color online).
(a) Low-energy electron diffraction patterns of Bi/InSb(001)-(1$\times$3) taken at room temperature (RT). 
(b) Surface and bulk Brillouin zones of Bi/InSb(001).
(c-e) ARPES results at RT measured with $h\nu$ = 21 eV.
(c) Constant energy contour around the Fermi level ($E_{\rm F}$).
The dashed lines indicate boundaries of the surface Brillouin zone (SBZ). $k_x$ and $k_y$ are defined parallel and perpendicular,  respectively, to [110].
ARPES intensities are integrated over 20 meV windows centered at $E_{\rm F}$.
(d) Momentum distribution curves in second SBZ.
(e) Band dispersion around $E_{\rm F}$ along $k_y$.
}
\end{figure}

\begin{figure}[p]
\includegraphics[width=80mm]{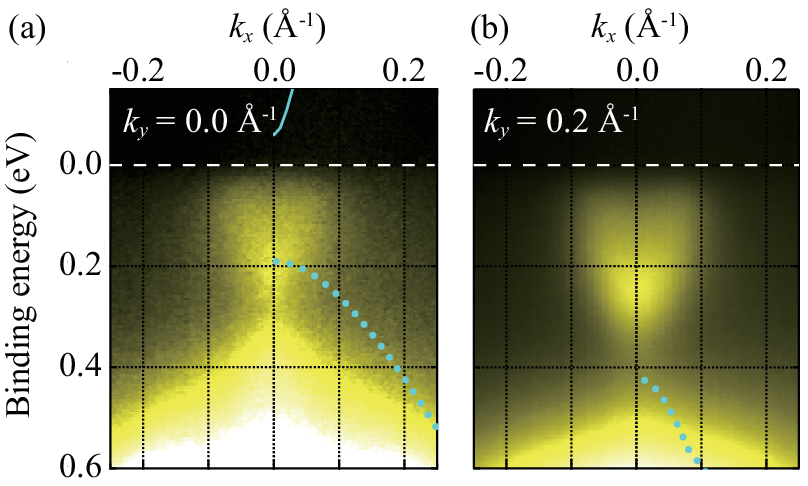}
\caption{\label{fig2} (Color online)
(a, b) Band dispersions along $k_y$ at 8 K with $h\nu$ = 15 eV.
The dotted (solid) curves represent the upper (lower) edges of projected bulk valence (conduction) bands.
}
\end{figure}

The Bi/InSb(001) surface was prepared by evaporating three monolayers of Bi on the clean surface of InSb(001) substrates ($n$-type, carrier concentration of 2$\times$10$^{14}$ cm$^{-3}$) and a subsequent flash by direct current heating up to 680 K for $\sim$10 seconds.
After the preparation, the $p$(1$\times$3) surface periodicity was observed by low-energy electron diffraction (Fig. 1(a)).
The detailed procedure on sample preparation and characterization, especially the comparison with the $c$(2$\times$6) surface reported in a previous work \cite{Laukkanen10}, is shown in the supplemental material (SM) \cite{SM}.

Figs. 1(c)–(e) indicate the surface electronic structure of Bi/InSb(001) measured with ARPES.
ARPES measurements were performed with a He lamp and synchrotron radiations at the CASSIOP\'EE beamline of synchrotron SOLEIL and BL7U of UVSOR-III.
The photon energies used ranged from 12 to 40.8 eV.
The photoelectron kinetic energy at $E_{\rm F}$ and the overall energy resolution of each ARPES setup were carefully calibrated by the Fermi edge of the photoelectron spectra from freshly evaporated gold on the sample after the measurement (see Fig. 3(a)).

Figure 1 (c) shows the Fermi contour (FC) around $E_{\rm F}$ as measured with linearly polarized photons.
$k_x$ and $k_y$ are defined to be parallel to [110] and [$\bar{1}$10], respectively.
The linear shape of the FC along $k_y$ clearly indicated a 1D metallic state on the surface. Such a 1D FC is consistent with the anisotropic (1$\times$3) surface periodicity indicated by LEED.
The dispersion of the 1D state did not change with the photon energy (shown in SM \cite{SM}), which indicates no dispersion along $k_z$ as would be expected for a surface state.
Figure 1 (c) also shows that there were no other metallic states on the surface.
Therefore, this FC provides a unique and highly anisotropic conduction path on the surface.
As shown in Fig. 1(d), the ARPES momentum distribution curves (MDCs) indicated that the SS was exactly 1D without any 2D distortion. 
MDCs were captured in the second SBZ because the SS dispersion became clearer there owing to the photoexitation matrix element effect.
The lack of 2D distortion can also be observed in the $E$-$k_y$ plot in Fig. 1(e).
The dotted curve is the upper edge of the projected bulk valence bands as calculated according to Ref. \cite{Chadi77}.
The bulk valence band maximum position was determined with the linear method (see SM for details \cite{SM}).

Note that the 1D FC appeared with a single domain, in contrast to the double-domain case of Au/Ge(001) \cite{Meyer14, Yaji15}, where the direction of the conduction path is still under debate because of the two equivalent domains rotated 90$^\circ$ with each other \cite{Park14, Nakatsuji12, Yaji15}.
This is because of the reduced symmetry of the zinc-blend crystal structure compared to group-IV semiconductors. 
Along the [001] direction, a III-V semiconductor lattice is stacked alternately with group III and V atoms (indium and antimony for InSb).
Hence, the equivalent plane appears at every two atomic steps on the (001) surface of such crystal. 
In the group IV semiconductor crystal, every atomic plane along [001] is equivalent and appears with 90$^\circ$ rotation at each atomic step to form the double-domain surface structure.
The 90$^\circ$ rotation at each atomic step also occurs for III-V semiconductors; the double atomic step results in a 180$^\circ$ rotation, so the steps are actually identical thanks to the two-fold rotation symmetry. 
Thus, the conduction path of the Bi/InSb(001) surface was clearly demonstrated to be parallel to [110].

Figure 2 shows the dispersion of the surface band on Bi/InSb(001).
The surface state formed a single parabolic band with its bottom at $k_x$ = 0 \AA$^{-1}$.
This band is the unique metallic feature on the surface.
This 1D metallic band crossed $E_{\rm F}$ only once on each side of the SBZ. 
At higher binding energies, the bottom of the surface-state band showed a small shift ($\sim$50 meV) between $k_y$ = 0.0 and 0.2 \AA$^{-1}$.
This would be due to an overlap of the bulk valence bands. 
The dotted (solid) curves in Fig. 2 are the upper (lower) edges of the projected bulk valence (conduction) bands and were calculated in the same manner as those in Fig. 1(e). 
The bottom of the SS overlapped the bulk valence bands at the center of the SBZ. 
Note that the bulk conduction band does not cross $E_{\rm F}$, as shown by the solid curve in Fig. 2(a).

\begin{figure}[p]
\includegraphics[width=80mm]{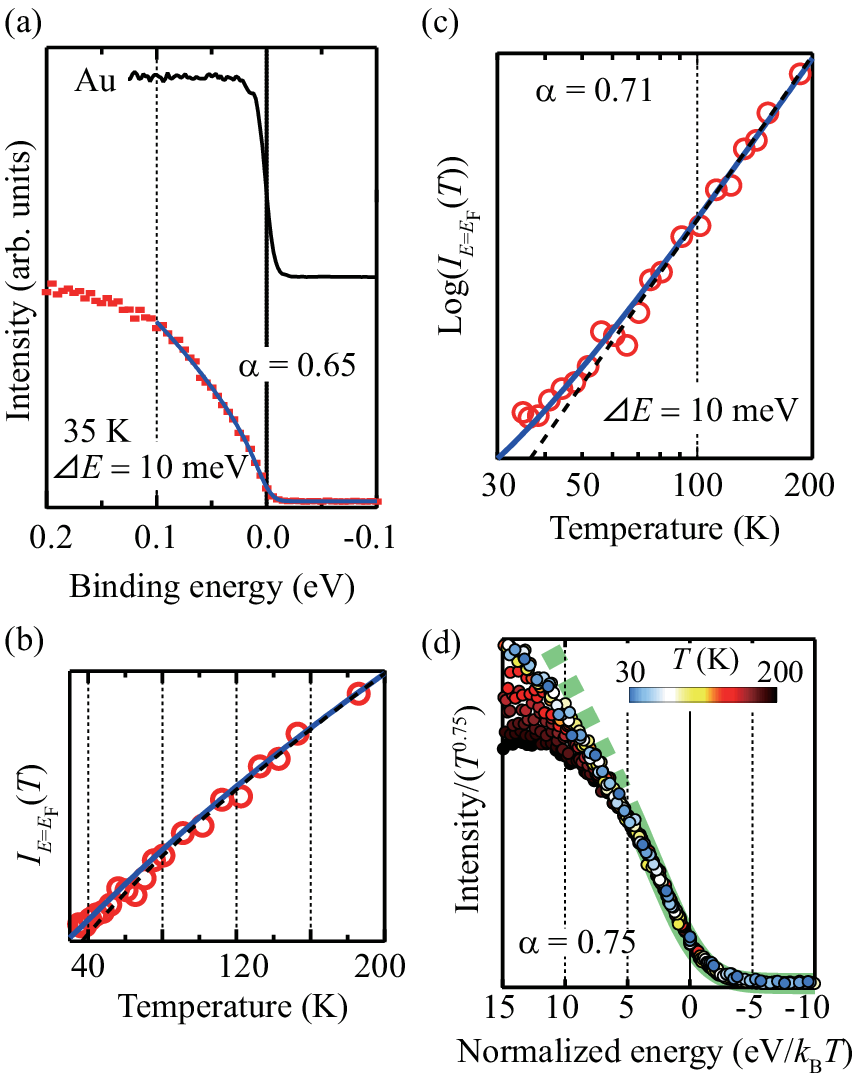}
\caption{\label{fig3} (Color online).
(a) Photoelectron spectrum (red markers) integrated along $k_x$ ($k_y$ = 0.25$\pm$0.05 \AA$^{-1}$).
The solid line between $\pm$ 0.1 eV is the fitting curve (see text for details).
(b) Temperature dependence of the photoelectron intensity at $E_{\rm F}$ ($I_{E=E_{\rm F}}(T)$). Each spectrum was normalized at the intensity at 0.2 eV below $E_{\rm F}$ in order to compensate the sample degradation.
The solid (blue) line is the fitting curve based on power-law scaling ($T^{0.71}$) convolved with $\Delta E$ (10 meV).
The dashed line is $T^{0.73}$ without broadening from $\Delta E$.
(c) The same as (b) plotted in a log scale.
(d) Universal scaling plot of the spectra at various temperatures, using a $T$-renormalized energy scale (eV/$k_{\rm B}T$). The thick dotted (green) line is from eq. (1) with the scaling factor $\alpha$ = 0.75.
}
\end{figure}

In order to examine the 1D nature of the surface state, we compared the angle-integrated photoelectron spectrum with that from ordinary metal (Au evaporated on the sample after the ARPES measurement), as shown in Fig. 3 (a).
The angle integration was performed at $k_y$ = 0.25$\pm$0.05 \AA$^{-1}$.
Figure 3 (a) clearly shows the suppression of the photoelectron intensity at $E_{\rm F}$ compared to that from Au, which suggests a deviation of the surface state from the Fermi-liquid framework.
We tried to fit the spectrum to the TLL spectral function considering the finite temperature, as given in some of the literature \cite{Schonhammer93, Orgad01}.
Based on ref. \cite{Schonhammer93}, the spectral function of the angle-integrated PES can be written as,
\begin{eqnarray}
I(\epsilon, T) \propto T^{\alpha}cosh\left(\frac{\epsilon}{2}\right )\left|\Gamma \left(\frac{1+\alpha}{2}+i\frac{\epsilon}{2\pi} \right) \right |^2f(\epsilon, T),
\end{eqnarray}
where $\epsilon$ = $E/k_{\rm B}T$ is the temperature-normalized energy, $\Gamma$ is the gamma function, and $f(\epsilon, T) = (e^{\epsilon}+1)^{-1}$ is the Fermi distribution function.
When convolved with the Gaussian instrumental energy resolution ($\Delta E$ = 10 meV), this function reproduced the spectra well at 35 K (Fig. 3 (a)) with $\alpha$ = 0.65 $\pm$ 0.05 in the low-energy region.
This function only fits the spectrum below 0.1 eV. 
This was because the TLL framework is only valid in the proximity of $E_{\rm F}$ (see SM for detailed analysis \cite{SM}).

The power-law scaling of a TLL should show an universal power index with various parameters.
Figs. 3 (b) and (c) show the temperature dependence of the photoelectron intensities at $E_{\rm F}$.
The intensity from 180 to 50 K decreased linearly with log $T$, which indicates power-law scaling of the spectral intensities, $I_{E=E_{\rm F}}(T) \propto T^{\alpha}$.
The slight change in the slope below 50 K was due to the finite $\Delta E$.
This broadening was already convolved in the fitting curve (blue solid line) to give a power index of $\alpha = 0.71\pm 0.10$, which is a value very close to that obtained by fitting the energy scale (Fig. 3 (a)).
Note that $I_{E=E_{\rm F}}(T)$ should be constant in the case of a Fermi-Dirac system.

Another method to examine universal power-law scaling is to normalize the spectra taken at various temperatures to $T^{\alpha}$ and plot them versus the temperature-normalized energy $\epsilon$, as shown in Fig. 3 (d).
With such a renormalization, all spectra with various $T$ become almost identical in the low-energy region, typically in $\epsilon \le$ 5.
For such identification, the scaling factor $\alpha$ = 0.75 $\pm$ 0.05 is required
(spectra renormalized with another $\alpha$ are shown in the SM \cite{SM}).
Because such a temperature normalization method can include the intensity not only at $E$=$E_{\rm F}$ but within the finite energy window, this renormalized plot would be the best way to estimate the power index from angle-integrated spectra taken at various temperatures.
All of these spectral features indicate that the 1D surface state on Bi/InSb(001) obeys an universal power-law scaling with $\alpha$ = 0.7$\pm$0.1, which agrees with that expected for a TLL.

\begin{figure}[p]
\includegraphics[width=80mm]{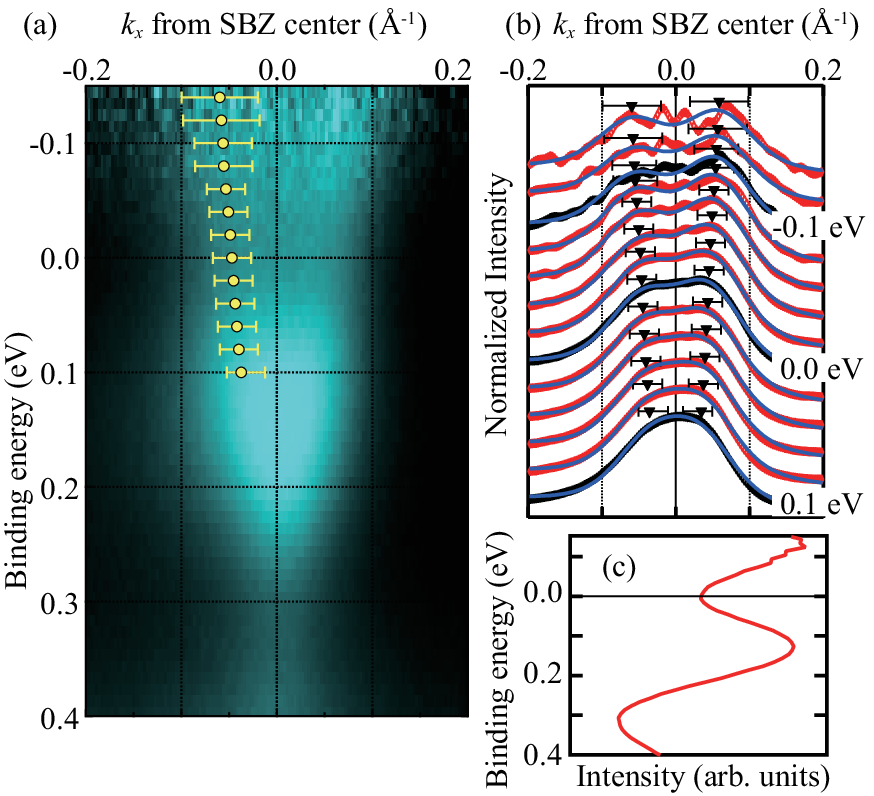}
\caption{\label{fig4} (Color online). 
(a) ARPES intensity plot along $k_x$ divided by the Fermi distribution function convolved with the instrumental resolution taken at $h\nu$ = 21 eV.
ARPES data were integrated along $k_y$ within $|k_y|<$ 0.3 \AA$^{-1}$ for better statistics.
(b) ARPES MDCs along $k_x$ within 20 meV energy window. The intensities of each curve were normalized by the peak height.
Solid lines are the fitting curves (see text for details).
(c) An angle-integrated EDC within $|k_x|<$ 0.3 \AA$^{-1}$ divided by the Fermi distribution function, similar to (a).
}
\end{figure}

To obtain further insight into the 1D SS, we looked into the detailed structure of ARPES spectra.
Figure 4 (a) shows the ARPES intensity plot at RT along $k_x$ divided by the Fermi distribution function convolved with the instrumental resolution.
For better statistics, the ARPES plot was integrated along $k_y$ within $|k_y|<$0.3 \AA$^{-1}$.
The momentum distribution curves (MDCs) are shown in Fig. 4 (b).
They are fitted well by a pair of Voigt peaks (Gaussian and Lorentzian widths are fixed for every peak) and linear backgrounds (solid lines in Fig. 4 (b)).
Weak residuals around $k_x$ = 0 \AA$^{-1}$ at the binding energies below -0.1 eV would be due to the bulk conduction bands.
From the MDCs, we obtained the peak positions at each binding energy and overlaid them on Fig. 4 (a).
They indicated the Fermi wavevector $|k_{\rm F}|$ to be 0.05 $\pm$0.02 \AA$^{-1}$.
This estimation is consistent with the ARPES data taken at low temperature shown in Fig. 2.
This shows the continuous, metallic dispersion of the 1D SS across $E_{\rm F}$.
Moreover, the ARPES intensity plot clearly shows that the spectral intensity of the 1D SS was  suppressed around $E_{\rm F}$, without breaking the continuity of the band dispersion.
Such suppression of the density of states (DOS) around $E_{\rm F}$ without any gap opening reflects the power-law scaling of the TLL.
This result excludes alternative possibilities without ambiguity.
Two different states below/above $E_{\rm F}$ cannot reproduce such a dispersion.
In addition, the angle-integrated spectrum showed a temperature-broadened power-law shape for the density of states around $E_{\rm F}$, as expected in theory \cite{Schonhammer93}.
At RT, it sometimes becomes difficult to determine the exact position of $E_{\rm F}$ because of the broad Fermi edge.
However, note that such ambiguity did not affect the DOS suppression obtained by ARPES: the intensity suppression appeared irrespective of slight artificial shifts in $E_{\rm F}$.

In summary, a new candidate for a surface TLL formed on Bi/InSb(001) has been discovered.
A surface state formed a 1D FC with almost no 2D undulation and stayed metallic down to 35 K.
This 1D FC is the only metallic feature of the sample and hence provides an unique 1D conduction path on the surface.
The photoelectron spectra in the proximity of $E_{\rm F}$ obeyed the power-law scaling as a function of binding energy and sample temperature with the same power index $\alpha$ = 0.7 $\pm$0.1.
Moreover, the ARPES plot above $E_{\rm F}$ obtained thanks to thermally broadened Fermi edge at RT indicated a 1D state with continuous metallic dispersion across $E_{\rm F}$ and power-law intensity suppression around $E_{\rm F}$.
These results strongly suggest that the 1D surface structure Bi/InSb(001) hosts a surface TLL.
The new 1D SS on Bi/InSb(001) would provide a fertile playground for further studies of low-dimensional physics.

We thank V. Meden for the helpful discussions.
We also acknowledge D. Ragonnet and F. Deschamps for their support during the experiments on the CASSIOP\'EE beamline at synchrotron SOLEIL.
Part of the ARPES experiments were performed under the UVSOR proposal No. 26-824 and Nanotechnology Platform Program of the Ministry of Education, Culture, Sports, Science and Technology (MEXT), Japan.
This work was also supported by the JSPS Grant-in-Aid for Research Activity Start-up (Grant No. 26887024).

\newpage

\renewcommand{\thefigure}{S\arabic{figure}}
\setcounter{figure}{0}

\section*{Supplementary material for: Surface Tomonaga-Luttinger liquid state on Bi/InSb(001)}

\subsection{Sample growth and characterization of Bi/InSb(001)-$p$(1$\times$3) and $c$(2$\times$6) surfaces}

InSb(001) substrates ($n$-type, carrier concentration of 2$\times$10$^{14}$ cm$^{-3}$) were cleaned by repeated cycles of sputtering and annealing up to 680 K until the $c$(8$\times$2) low-energy electron diffraction (LEED) pattern was observed.
Then, a nominal 3 monolayers (ML) of Bi were evaporated at room temperature: 1 ML is defined as the atom density of bulk-truncated InSb(001) and the coverage was estimated by using a quartz microbalance.
A subsequent flash by direct current heating up to 680 K for $\sim$10 seconds produced the $p$(1$\times$3) surface, as indicated by the sharp and low-background LEED pattern shown in Fig. 1 (a) of the main text.
As shown in Fig. S1 (a), the excess Bi was removed after the flash.
From the height of Bi 5$d$ with linear-background subtraction, the coverage of Bi on the (1$\times$3) surface was $\sim$1.5 ML.

Sometimes, we could find faint fractional spots showing the $c$(2$\times$6) periodicity on the flashed surface, which suggested a small fraction of the coexisting area with the $c$(2$\times$6) periodicity, as shown in Fig. S1 (b).
The previous work \cite{Laukkanen10} reported a Bi/InSb(001)-$c$(2$\times$6) surface prepared by 2 ML of Bi evaporation followed by annealing at $\sim$ 470 K for 2 hours.
This surface was reported to be metallic based on the angle-integrated photoelectron spectrum taken at room temperature (RT).
For comparison, we prepared the $c$(2$\times$6) surface according to Ref. \cite{Laukkanen10}.
The LEED pattern showed the $c$(2$\times$6) periodicity (see Fig. S1 (c)).
As shown in Fig. S2 (a), the Fermi contour measured by ARPES did not show any one-dimensional SS.
Instead, there was a two-dimensional metallic state.
The angle-integrated spectra in Fig. S2 (b) qualitatively agrees with that in the previous work.

Based on these results, we concluded that the 1D SS observed in this work comes from the flashed (1$\times$3) surface and that the $c$(2$\times$6) area does not contribute at all to the 1D states.

\subsection{ARPES images taken with different photon energies}
Figure S3 shows the ARPES images taken with various photon energies.
The parabolic state below the Fermi level showed no $k_z$ dispersion; thus, it is the surface state.

\subsection{Bulk valence band maximum (VBM)}
Figure S4 shows the angle-integrated photoelectron spectra taken with four different photon energies.
We determined the energy position of VBM to be the intersection between the linear fits of the background and the linear portion of the valence band leading edge \cite{VBM}.
Based on the so-called linear method, the valence band maximum was estimated to be 0.19$\pm$0.01 eV.

\subsection{Fitting curve for angle-integrated photoelectron spectra}
We checked the dependence of the power index to the energy window as shown in Fig. S5.
Fitting was performed based on Eq. (1) in the main text.
Below 0.1 eV, the power index was almost the same and the fitting curve agreed well with the spectrum.
With higher binding energies, the power index tended to become smaller and Eq. (1) failed to reproduce the experimental data.
This must be because the TLL framework is only valid in the low-energy region.
In addition, this deviation from the theoretical curve would be because the bottom of the SS was $\sim$ 0.25 eV: around the bottom of the surface-state band, the dispersion is no longer linear.

\subsection{Renormalized photoelectron spectra with various scaling factors}
Figure S6 shows the universal scaling plot of the angle-integrated photoelectron spectra with different scaling factors $\alpha$.
Fitting of the spectra taken at various temperatures to the universal curve was optimized at $\alpha$ = 0.75.

\begin{figure}
\includegraphics[width=80mm]{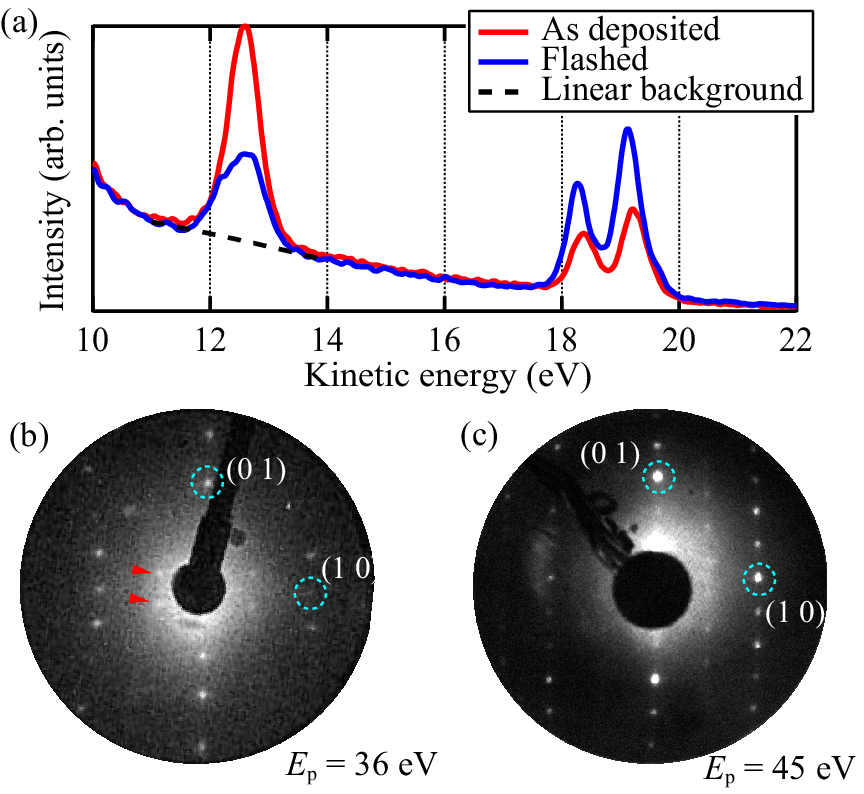}
\caption{\label{fig1s}
(a) Core-level spectrum for the as-deposited and flashed Bi/InSb(001) surfaces taken at $h\nu$ = 40.8 eV (He II).
(b) LEED pattern of a flashed Bi/InSb(001) surface with coexisting $c$(2$\times$6) spots (indicated by red markers). The pattern is distorted because of the flat MCP screen.
(c) LEED pattern of the Bi/InSb(001)-$c$(2$\times$6) surface.
}
\end{figure}

\begin{figure}
\includegraphics[width=80mm]{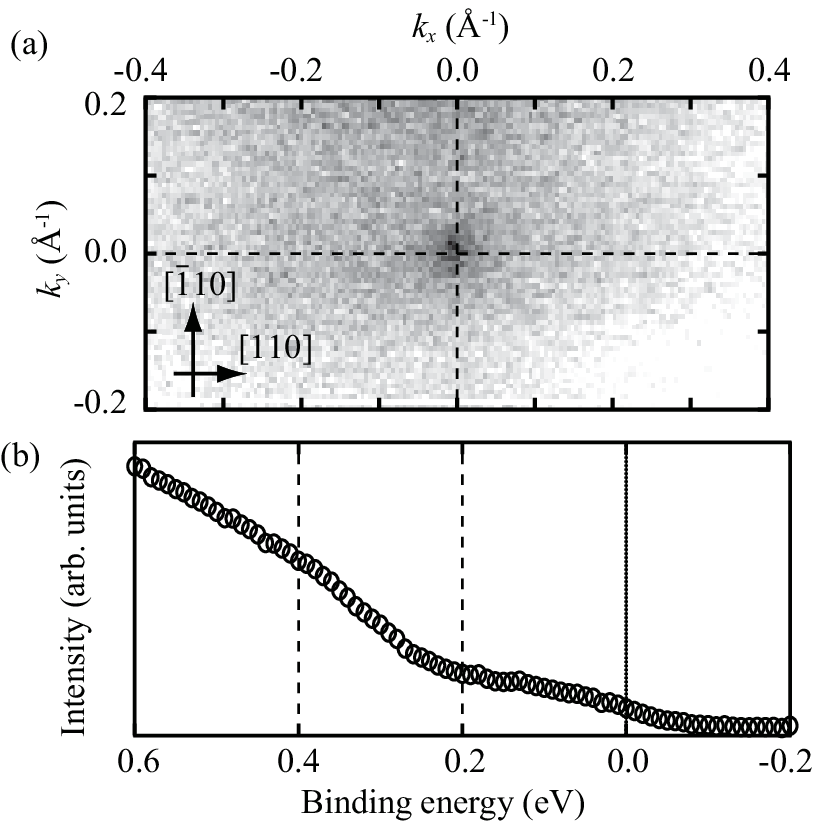}
\caption{\label{fig2s}
ARPES results of the Bi/InSb(001)-$c$(2$\times$6) surface with $h\nu$ = 19 eV at room temperature.
(a) Fermi contour.
(b) Angle-integrated spectrum $k_x$ ($k_y$ = 0 \AA$^{-1}$).
}
\end{figure}

\begin{figure}
\includegraphics[width=80mm]{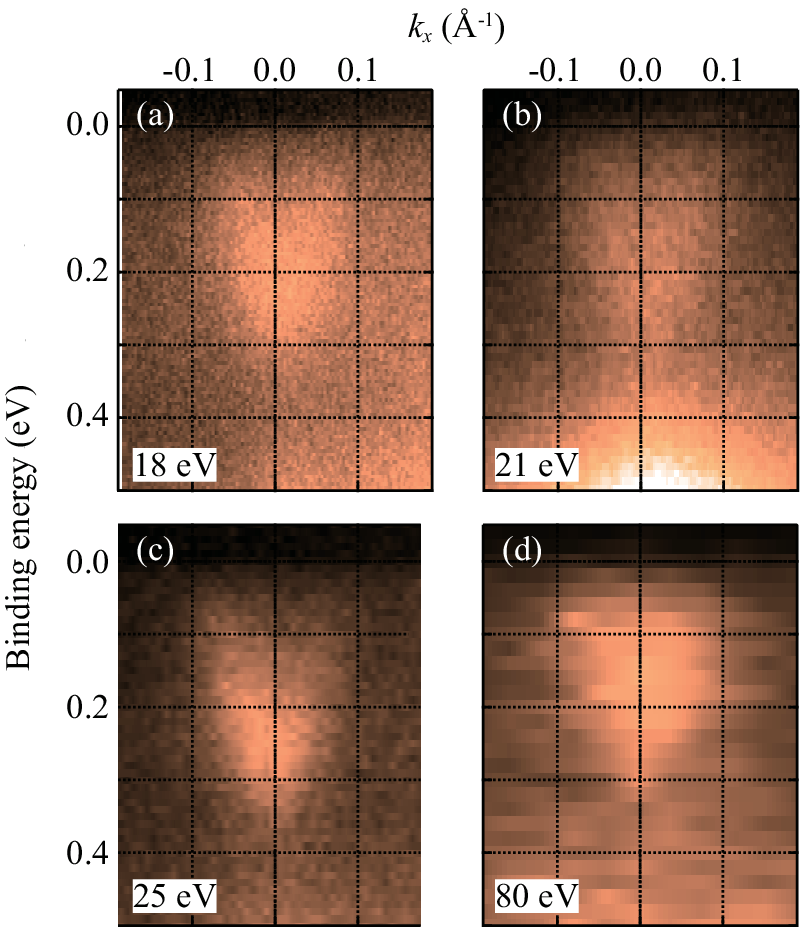}
\caption{\label{fig3s}
ARPES intensity maps of Bi/InSb(001) along $k_x$ ($k_y$ = 0.2 \AA$^{-1}$) with various photon energies taken at 8 K.
}
\end{figure}

\begin{figure}
\includegraphics[width=80mm]{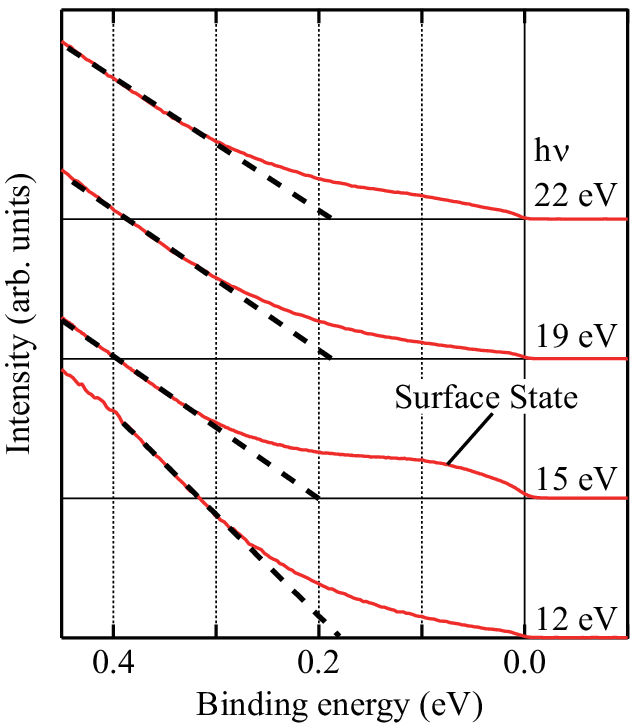}
\caption{\label{fig4s}
Angle-integrated photoelectron spectrum along $k_x$ ($k_y$ = 0 \AA$^{-1}$) measured at 13 K. 
The dashed lines are linear fits of the leading edges from the bulk valence bands.
}
\end{figure}

\begin{figure}
\includegraphics[width=70mm]{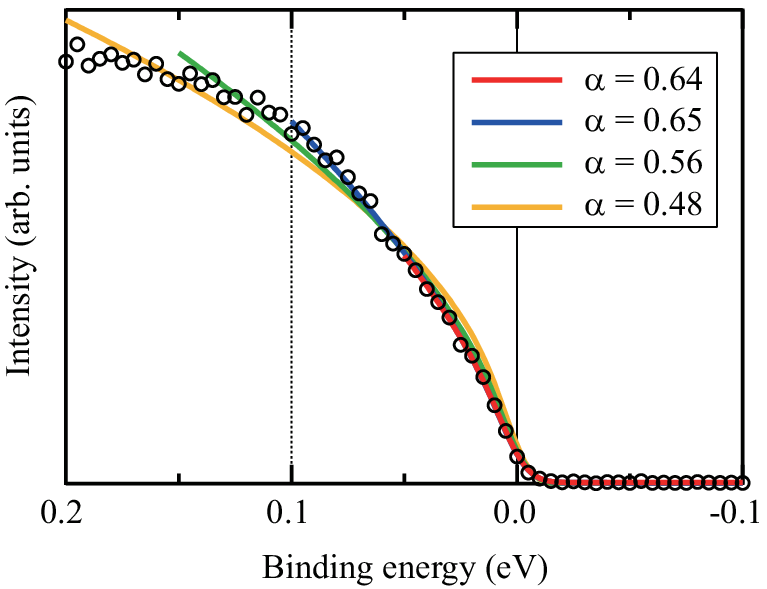}
\caption{\label{fig5s}
Angle-integrated photoelectron spectrum (same data as figure 3 (a)) fitted with various energy ranges.
}
\end{figure}

\begin{figure*}
\includegraphics[width=140mm]{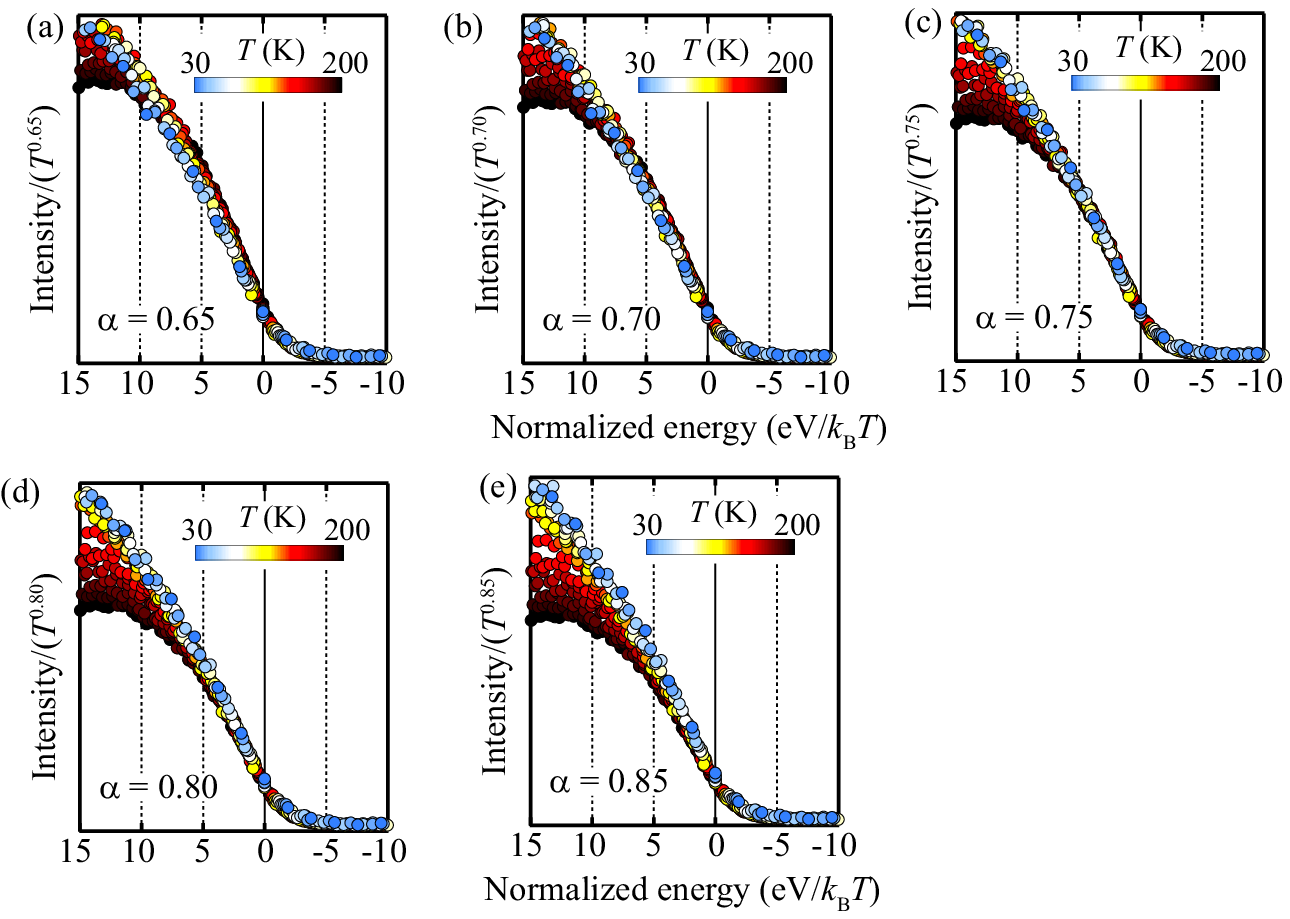}
\caption{\label{fig6s}
Universal scaling plot of the spectra at various temperatures with various scaling factors $\alpha$, using a $T$-renormalized energy scale (eV/$k_{\rm B}T$).
(a) $\alpha$ = 0.65, (b) $\alpha$ = 0.70, (c) $\alpha$ = 0.75, (d) $\alpha$ = 0.80, and (e) $\alpha$ = 0.85.
}
\end{figure*}

\end{document}